\documentclass[twocolumn,amsmath,amssymb,pra]{revtex4-1}
\usepackage{graphicx,amsfonts}
\DeclareFontFamily{U}{wncy}{}
\DeclareFontShape{U}{wncy}{m}{n}{<->wncyr10}{}
\DeclareSymbolFont{mcy}{U}{wncy}{m}{n}
\DeclareMathSymbol{\Sh}{\mathord}{mcy}{"58} 

\begin{document}
\title{Uncertainty in the context of multislit interferometry}
\author{Johannes CG Biniok}
\email{jcgb500@york.ac.uk}
\affiliation{University of York, York YO10 5DD, UK} 
\author{Paul Busch}
\email{paul.busch@york.ac.uk}
\affiliation{University of York, York YO10 5DD, UK}
\author{Jukka Kiukas}
\email{jukka.kiukas@nottingham.ac.uk}
\affiliation{University of Nottingham, Nottingham NG7 2RD, UK} 
\date{\today}
\begin{abstract}A pair of uncertainty relations relevant for quantum states of multislit interferometry is derived, based on the mutually commuting ``modular'' position and momentum operators and their complementary counterparts, originally introduced by Aharonov co-workers. We provide a precise argument as to why these relations are superior to the standard Heisenberg uncertainty relation at expressing the complementarity between spatial localisation and the appearance of fringes. We further support the argument with explicit computations involving wavefunctions specifically tailored to the interference setup. Conceptually developing the idea of Aharonov co-workers, we show how the modular momentum should reflect the given experimental setup, yielding a refined observable that accurately captures the fine structure of the interference pattern.\end{abstract}\maketitle

\newcommand{\sdev}[1]{\Delta{\big(#1\big)}} % for the standard deviation
\newcommand{\ft}[1]{\widehat{#1}} % formal Fourier transform
\newcommand{\Cos}[1]{\cos\left(#1\right)} % cos with brackets
\newcommand{\sinc}[1]{\mathrm{sinc}{\left(#1\right)}} % sinc function
\newcommand{\bracks}[1]{\left(#1\right)} % for brackets
\newcommand{\Int}[4]{\int_{#1}^{#2}{#3}{\,\mathrm{d} #4}} % for integrals
\newcommand{\MOD}[1]{#1_{\mathrm{mod}}} % APP's Pmod and Qmod
\newcommand{\sbracks}[1]{\left[#1\right]} % square brackets
\newcommand{\com}[2]{\left[#1,#2\right]} % commutator
\newcommand{\dom}[1]{\mathcal{D}(#1)}
\newcommand{\expo}[1]{\exp{\bracks{#1}}} % exp(#1)
\newcommand{\ImUnit}{\mathrm{i}} % for the imaginary unit i
\newcommand{\Euler}{\mathrm{e}} % for the exponential function
\newcommand{\Dcomb}[1]{{\Sh}_{#1}} % for the Dirac comb
\newcommand{\modsquare}[1]{|{#1}|^2} % for the mod square

\section{Introduction} Heisenberg's principle of uncertainty expresses the notion of a fundamental limitation of precise values for a pair of complementary observables for any  quantum state. Mathematically the uncertainty principle is often expressed as a tradeoff between the standard deviations of the relevant observables. In its most common form, the so-called Heisenberg uncertainty relation, a tradeoff is expressed between the standard deviations of the position $Q$ and the momentum $P$ of a single non-relativistic particle. Using $\sdev{Q,\Psi}$ to denote the standard deviation of $Q$ in state $\Psi$ and similarly $\sdev{P,\Psi}$ to denote the standard deviation of $P$, we have
\begin{equation}\label{U}
	\sdev{Q,\Psi}\,\sdev{P,{\Psi}}\geq\frac{1}{2},
\end{equation}
in units where $\hbar=1$. In the context of multislit interferometry though, it is clear already from an intuitive point of view that the relevant complementary observables are not exactly position and momentum. Indeed, position should be replaced by ``which slit" information, and momentum with fringe width. The choice for a mathematical representation of the former is clear (simply a coarse-grained position), but fringe width is less obvious. As is well known -- {see, for instance, Ref.~\onlinecite{UH}} -- and also argued here, the standard deviation of momentum cannot describe fringe width, rendering the Heisenberg relation unsuitable for expressing complementarity in the interferometric context. This may seem particularly surprising considering the important role played by the uncertainty relation \eqref{U} in the historic Bohr-Einstein debate, which was concerned, {\em inter alia}, with the complementarity of path information and interference contrast. 

It is not that the standard deviation as such is a poor measure though; it is the particular combination of standard deviation and the momentum operator $P$ that is problematic. 
A more suitable expression of the uncertainty principle may be found by using observables that take into account the periodic nature of the experimental setup. Mathematically, the idea is to modify the pair $(Q,P)$ by making $Q$ discrete and $P$ periodic. When properly adjusted, the resulting pair will still have a ``canonical" nature that leads to an uncertainty relation. We are going to derive commutation relations for such operators, which are formally similar to the standard relation \eqref{U}. The derived relations, however, will be valid only for wavefunctions of a subset specific to the interferometric setup under consideration. A proper understanding of this limited validity of the uncertainty relation requires careful consideration of the mathematical subtleties of the problem, in particular of domain questions.

The adaptation of the observables $Q$ and $P$ to the interferometric context is due to Aharonov {\em et al.}, who also provided a heuristic argument that the uncertainty relations discussed here should exist \cite{APP}. However, their work was never developed beyond invoking an analogy to angle and angular momentum. Most importantly, the fact that the relations are only valid for specific wavefunctions was never made clear. This is perhaps due to insufficient mathematical development of the problem in their work and in related publications \cite{APP, AR2005,Tollaksen2010}. It was not until the work of Gneiting and Hornberger of 2011 that correct commutation relations were stated \cite{Gneiting2011}, but the discussion there is mainly formal and with a different focus. In conclusion, a thorough analysis seems in order. We present here, to our knowledge for the first time, a precise derivation and discussion of these uncertainty relations. Furthermore, we address limitations and benefits of the uncertainty relations, and discuss an application to uniformly illuminated apertures and the asymptotic behavior of the uncertainty product in this case. We found that further conceptual development of the idea of Aharonov \emph{et al.} was necessary for describing the fine structure of quantum states as prepared in a given interferometry experiment. Precisely resolving the fringe structure of the interference wavefunction is only possible with an observable that is adapted specifically to the particular experimental setup. We thereby obtain the expected asymptotic behavior and find that the uncertainty product converges to a finite value in the limit of infinitely many illuminated slits. The underlying argument follows from considerations involving a product form of a certain physically motivated set of interference wavefunctions. This product form, although mathematically elementary, provides interesting insight into the structure of interference wavefunctions and apparently remained unnoticed. 

Throughout we will emphasize the interplay between physics and mathematics leading to a derivation that is mathematically deceptively simple and physically insightful, although occasionally subtle on both accounts. From a more practical point of view, we also provide a simple way of computing the uncertainty products, based on the aforementioned product form of the wavefunctions. (Otherwise rather involved integrations would be required.) 

\section{Observables of multislit interferometry} 
\begin{figure}\centering
	\includegraphics[width=0.5\textwidth]{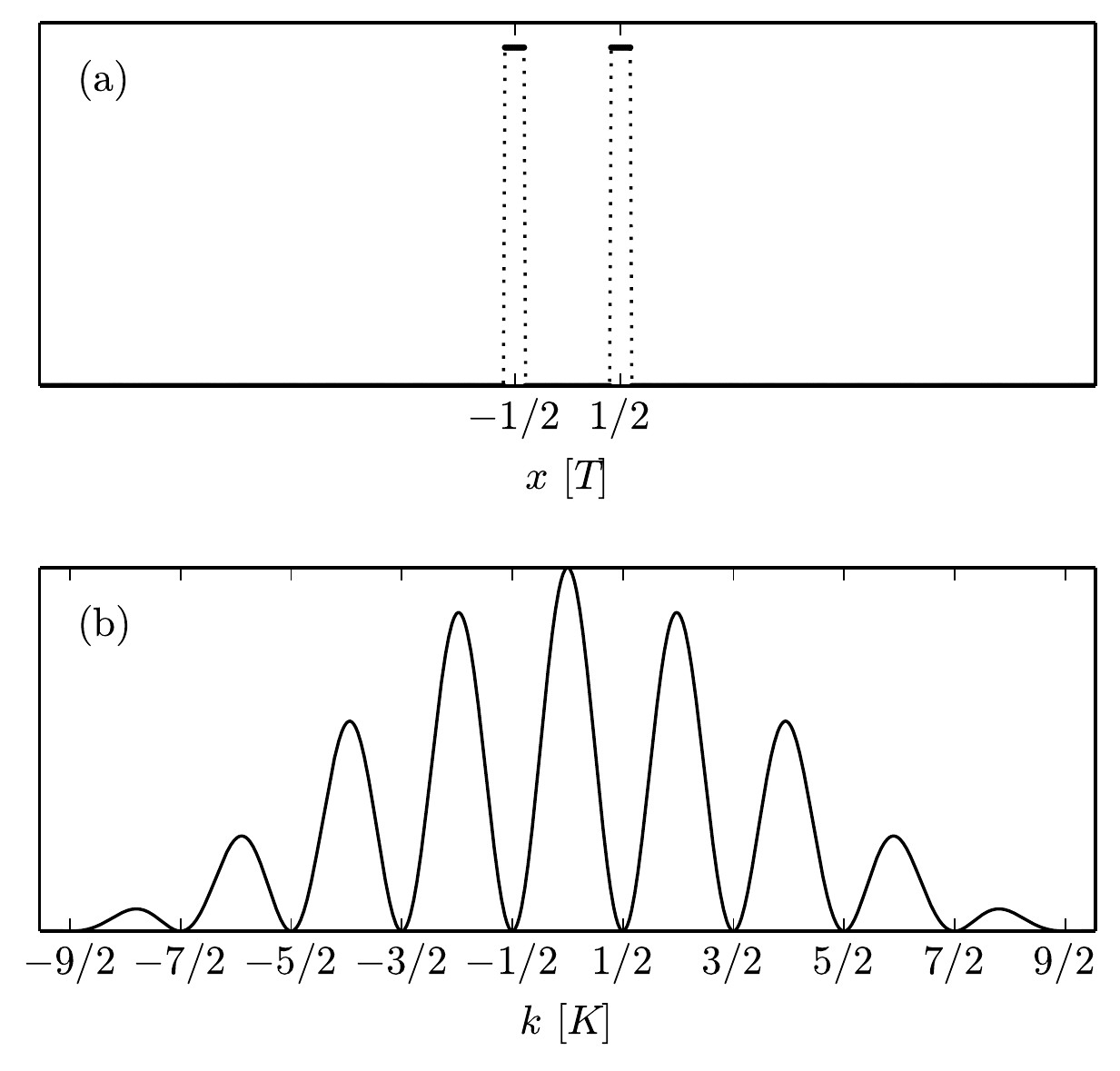}
	\caption{Intensity profiles associated with the double-slit state $\psi_2$ are depicted, position space in $\mathrm{(a)}$ and in $\mathrm{(b)}$ momentum space. Note in particular that $\ft{\psi}_2(k)$ vanishes at $k=(j+1/2)K$, with integer $j$.}
	\label{F1}
\end{figure}

The traditional approach, employing relation \eqref{U}, fails at quantifying the uncertainty of a double-slit superposition state. Assuming rectangular slits of width $a$ and a wavefunction $\Psi(x)$ of constant real-valued amplitude passing through an array of slits, the intensity profile of the fringe pattern with slits at locations $\pm T/2$ is given by the Fourier transform
\begin{equation}\label{ds}
	\left|\ft{\psi}_{2}(k)\right|^2=\frac{a}{\pi}\,\sinc{\frac{a}{2}\,k}^2\Cos{\frac{T}{2}\,k}^2,
\end{equation}
where the {\em cosine} describes the fringes, and the {\em sinc} describes an envelope. For an illustration, see Fig.~\ref{F1}. As $\sdev{P,\psi_{2}}$ diverges, relation \eqref{U} provides no information. The presence of fringes, or lack thereof, has no impact on the result. This is particularly apparent when considering the single-slit state $\psi_1$, for which again $\sdev{P,\psi_{1}}=\infty$.
The root of this problem is the combination of standard deviation and operator $P$. For instance, the moments of $P$ are insensitive to the relative phase between two path states, or even to the absence of a phase relation in the case of a mixed state. This led Aharonov {\em et al.} to instead consider unitary shift operators for a description of interference, as these can be used to create overlap and thus establish sensitivity to relative phase. They then proposed a decomposition of the noncommuting operators $Q$ and $P$ into commuting parts $\MOD{Q}$ and $\MOD{P}$ (periodic) and noncommuting parts $Q_T$ and $P_K$, and presented a heuristic argument that an uncertainty relation of the same form as the Heisenberg relation \eqref{U} should exist.

The following observation illustrates this idea: $Q$ being the shift generator for quantum states in momentum space and $P$ being the shift generator for position space, the unitary shift operators considered by Aharonov {\em et al.} are, in fact, identical to the shift operators in Weyl's commutation relation
\begin{equation}
	\Euler^{\ImUnit\,p\,Q}\,\Euler^{\ImUnit\,q\,P}=\Euler^{-\ImUnit pq}\,\Euler^{\ImUnit\,q\,P}\,\Euler^{\ImUnit\,p\,Q}.
\end{equation}
It follows immediately that the operators $\Euler^{\ImUnit\,p\,Q}$ and $\Euler^{\ImUnit\,q\,P}$ commute for $pq=2\pi n$, with $n\in\mathbb{N}$; equivalently for the relative periods $T$ and $2\pi/(nT)$. For the present, we restrict our attention to $n=1$, and define
\begin{equation}\label{K}
	2\pi/T=:K.
\end{equation}
This observation suggests a decomposition of $Q$ into a $T$-periodic part and a remainder, and $P$ into a $K$-periodic part with remainder. More precisely,
\begin{align}
Q&=\MOD{Q}+Q_T,\label{decQ}\\
P&=\MOD{P}\,+P_K,\label{decP}
\end{align}
yielding a pair of commuting operators
\begin{equation}
	[\MOD{Q}, \MOD{P}]=0. \label{MV}
\end{equation} 
The subscript ``mod'' was chosen to reflect the terminology of Aharonov {\em et al.}, who refer to these observables as ``modular variables.'' We require the following definitions:
\begin{align}
	\MOD{Q}&=Q\mod{T},\label{Qmod}\\
	\MOD{P}&=(P+K/2\mod{K}) - K/2.\label{Pmod}
\end{align}
Note that $\MOD{P}$ is shifted by half the fringe separation in order to avoid overlap of the fringes with the discontinuous part of $\MOD{P}$. This shift is crucial for avoiding anomalous behavior of $\sdev{\MOD{P},\psi_2}$, as will become evident shortly; Aharonov {\em et al.} appear to have neglected it \cite{APP,AR2005,Tollaksen2010}. The definitions of the operators $Q_T$ and $P_K$ follow from $\eqref{decQ}$ and $\eqref{decP}$.
\begin{figure}
	\includegraphics[width=0.49\textwidth]{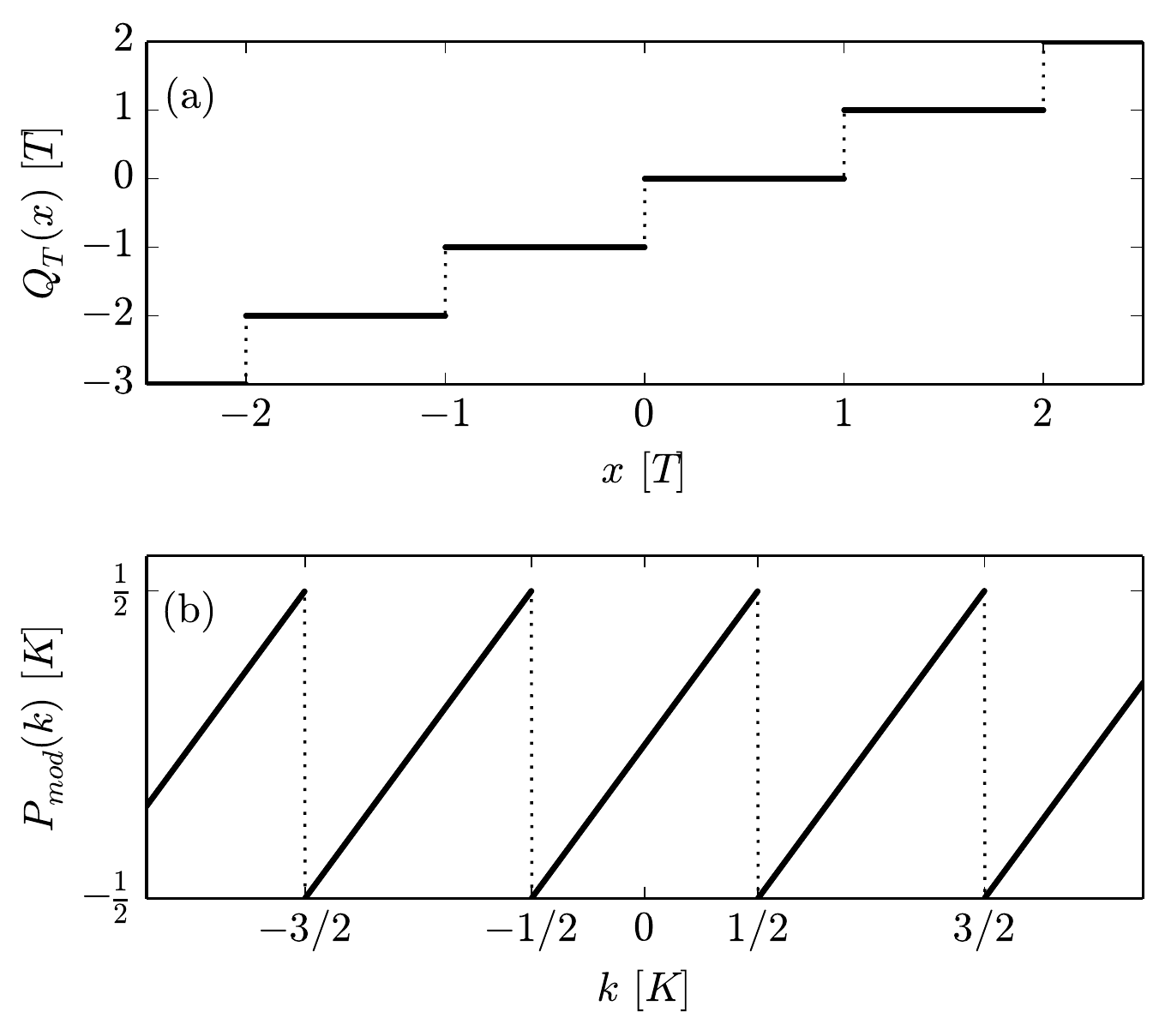}
	\caption{In (a) $Q_T$ is depicted in position space; it is defined indirectly through relations \eqref{Qmod} and \eqref{decQ}, whereas (b) illustrates $\MOD{P}$ in momentum space, as defined in \eqref{Pmod}.
	}
	\label{F0}
\end{figure}
$Q_T$ corresponds to a discretized position observable, while $P_K$ is a discretized momentum observable. Illustrations of $Q_T$ and $\MOD{P}$ are displayed in Fig.~\ref{F0}.
 
\section{Uncertainty relations adapted to multislit interferometry}\label{APPsec} Conceptually the periodic quantity $\MOD{P}$ seems appropriate for measuring the fine structure of the momentum distribution. Similarly, $Q_T$ appears suitable for measuring the localisation property as it corresponds to a coarse position observable -- the spatial localisation should not critically depend on the exact shape of the slits. Moreover, $[\MOD{P},\MOD{Q}]=0$ suggests that the canonical nature of the pair $(Q,P)$ is ``transferred" to the pairs $(Q_T,\MOD P)$ and $(\MOD Q,P_K)$, the only problem being that $Q_T$ and $P_T$ have discrete spectrum and so cannot be canonical operators in the strict sense.
In fact, it is possible to obtain the following two commutators for general quantum states \cite{Gneiting2011,BBK} 
\begin{align}
	&\com{Q_T}{\MOD{P}}=\ImUnit\,{\bf 1}-\ImUnit\,K\,\Dcomb{K}((\cdot)-K/2),\label{Jukka1}
	\\&\com{\MOD{Q}}{P_K}=\ImUnit\,{\bf 1}-\ImUnit\,T\,\Dcomb{T}.\label{Jukka2}
\end{align}
Here we have introduced the Dirac comb $\Dcomb{T}$, which we denote using the Cyrillic letter `sha' as is occasionally done in electrical engineering. The Dirac comb (or Shah function) is defined as 
\[\Dcomb{T}(x)=\sum_{j=-\infty}^{\infty}{\delta(x-jT)},\]
consisting of periodically spaced delta-distributions $\delta$, with $T$ denoting the spacing. The two relations \eqref{Jukka1} and \eqref{Jukka2}, are unsuitable for obtaining uncertainty relations resembling \eqref{U}, because of the state-dependent term. Simplifying these commutators to a canonical form is only possible if the wavefunction vanishes at the locations of the delta peaks -- this leads to the aforementioned restriction on the validity of the associated uncertainty relations.

A rigorous derivation of the two commutation relations \eqref{Jukka1} and \eqref{Jukka2} is rather long and technical; it will be provided in Ref.~\onlinecite{BBK}. It proceeds by making precise the analogy with the angular momentum-angle case alluded to by Aharonov {\em et al}. In fact, relations of the form \eqref{Jukka1}, \eqref{Jukka2} have been known for the angular momentum and angle pair since the 1960s \cite{JL63,SG64,CN}.

Here we present an alternative argument that immediately leads to the desired commutators by exploiting the properties of the relevant quantum states from the beginning. For the detailed discussion we focus on the pair $Q_T$ and $\MOD{P}$ as it is more natural to considerations in multislit interferometry; an analogous argument holds for $P_K$ and $\MOD{Q}$. In order to determine the commutator
\begin{equation}
	\com{Q_T}{\MOD{P}}\psi=\bracks{Q_T\MOD{P}-\MOD{P}Q_T}\psi
\end{equation} it is necessary to ensure that 
\begin{align}
	&\psi\in\mathcal{D}(Q_T)=\mathcal{D}(Q),\label{c0}\\
	&\MOD{P}\,\psi\in\mathcal{D}(Q_T)\label{c1}.
\end{align}
Here $\mathcal D$ denotes the domain of the indicated operator. Recall that the \emph{domain} of an operator is a subspace of square integrable wavefunctions (elements of the Hilbert space $L^2(\mathbb R)$), which, upon application of the operator, yield wavefunctions that are still square integrable. Since $\MOD{P}$ is bounded, its domain is the whole Hilbert space and hence does not lead to restrictions.

First, note that the domains of $Q$ and of $Q_T$ are equal, because these two operators differ by the bounded $\MOD{Q}$. Second, noting that  $Q$ acts as a differentiation operator in momentum space, a wavefunction in its domain is required to be (even absolutely) continuous. While \eqref{c0} thus amounts to the standard continuity assumption, the relation \eqref{c1} is peculiar to the present setup and requires more care. Since
\begin{align}\label{modP1}
\MOD{P}\,\ft{\psi}(k)&=(k-jK)\ft{\psi}(k) \nonumber\\
&\text{ for } k\in \left(\bigl(j-\tfrac12\bigr)K,\bigl(j+\tfrac12\bigr)K\right]
\end{align}
and $j\in \mathbb Z$, this function is discontinuous at $k=jK+K/2$, \emph{unless} $\ft{\psi}(k)$ vanishes at these points. This gives the aforementioned restriction on the wavefunction, explicitly:
\begin{equation}\label{con1}
	\ft{\psi}((j+1/2)K)=0 \text{ for each } j\in \mathbb Z.
\end{equation}
Since $\ft{\psi}$ is absolutely continuous by \eqref{c0}, this restriction is also sufficient for the commutator $\com{Q_T}{\MOD{P}}\psi$ to be defined.

Note that if $\MOD{P}\ft{\psi}(k)$ were not continuous, the derivative would not approach a finite limit value at the point of discontinuity; more precisely, we could describe the point of discontinuity by a step function, whose (distributional) derivative is a delta function -- this line of reasoning would eventually lead to the state-dependent correction terms in \eqref{Jukka1} and \eqref{Jukka2}.

Wavefunctions that naturally appear in the interferometric context typically have \emph{nodes} (i.e. zeros) periodically. As is evident from comparing Figs.~\ref{F1} and \ref{F0} and discussed in more detail below, our specific choice for $\MOD{P}$ in \eqref{Pmod} has its discontinuity points aligned with the nodes of the particular wavefunctions considered here. We emphasize once more that a more general wavefunction $\ft{\psi}(k)$ with periodic but non-vanishing values at the discontinuity points of $\MOD{P}$ is unsuitable because of boundary effects that lead to a state-dependent commutator.

Condition \eqref{con1} suggests a decomposition of the (dense) subspace of the admissible wavefunctions into a direct sum of subspaces
\begin{align*}
	\mathcal{D}_j&=\big\{\ft{\psi}\in L^2(jK+[-K/2,K/2])\mid \\
	&\ft{\psi}(jK-K/2)=\ft{\psi}(jK+K/2)=0\big\},
\end{align*}
where $j\in \mathbb Z$. Note that a restriction to any of the subspaces $\mathcal{D}_j$ corresponds to a quantum particle confined to a (`momentum') box, carefully discussed in the work of Bonneau {\em et al.} \cite{BFV}; below we point out parallels.

With the restriction \eqref{con1} on the wavefunction, $\MOD{P}$ corresponds to $P$ up to a constant, on each interval $jK+[-K/2,K/2]$, see Eq.~\eqref{modP1}. This enables us to perform the following formal manipulations in order to obtain the commutator
\begin{align}
	\com{Q_T}{\MOD{P}}&=\com{Q-\MOD{Q}}{\MOD{P}}
	\\&=\com{Q}{\MOD{P}}
	\\&=\com{Q}{P}&\text{on each $\mathcal{D}_j$}
	\\&=\ImUnit
\end{align}
While the algebraic manipulations are trivial, the penultimate expression may only be obtained by way of the domain considerations above. Hence, for each wavefunction in the dense subspace given by \eqref{con1} (together with the domain conditions \eqref{c0}, \eqref{c1}), we have
\begin{equation}
	\com{Q_T}{\MOD{P}}\psi=\ImUnit\,\psi.\label{com1}
\end{equation}
By means of the Robertson relation \cite{Robertson} for operators $A$ and $B$
\begin{equation}
	\sdev{A,\psi}\;\sdev{B,\psi}\geq\frac{1}{2}\bigl|\left\langle\psi|\com{A}{B}\psi\right\rangle\bigr|,
\end{equation}
the desired uncertainty relation now follows immediately:
\begin{equation}
	\sdev{Q_T,\psi}\;\sdev{\MOD{P},{\psi}}\geq\frac{1}{2}.\label{APP1}\end{equation}
This is the central result of the present investigation.
For completeness, we point out that this discussion proceeds analogously for wavefunctions restricted similarly in position space, yielding 
\begin{equation}
	\com{\MOD{Q}}{P_K}{\eta}=\text{i}\,{\eta},\label{com2}
\end{equation}
and ultimately
\begin{equation}
	\sdev{\MOD{Q},\eta}\;\sdev{P_K,{\eta}}\geq\frac{1}{2},\label{APP2}
\end{equation}
for wavefunctions $\eta$ from a suitably restricted dense subspace.

\section{Remarks} This section contains a number of conceptual and technical points and some critical observations on the work of Aharonov {\em et al}. \cite{APP,AR2005,Tollaksen2010}.

While the pairs of operators appearing in \eqref{APP1} and \eqref{APP2} are more appropriate for multislit interferometry than \eqref{U}, it must be stressed that these inequalities are valid only for quantum states $\psi$ and $\eta$, respectively, which vanish at the points of discontinuity of $\MOD{P}$ and $\MOD{Q}$. Notable exceptions are the eigenstates of $Q_T$ and $P_K$. In particular, the uncertainty relation \eqref{APP1} is inappropriate for a description of single-slit states. This point is particularly intriguing, because it is the adaptation of the observables to multislit interferometry that rules out the description of single-slit states.

The heuristic argument provided by Aharonov \textit{et al.} refers to the analogy between the pair $(Q_T,\MOD{P})$ and the pair of angular momentum and angle operators, the latter being understood as in the review of Carruthers and Nieto, Ref.~\onlinecite{CN}. The analogy can be made precise by observing that the restriction of the Weyl commutation relations to a discrete set of position variables and periodic set of momentum variables defines a representation of the Weyl relations on the group $\mathbb Z\times\mathbb T$, where $\mathbb T$ is the circle group. This representation is reducible, corresponding to the fact that the eigenspaces of $Q_T$ are infinite-dimensional, and can be decomposed into a direct sum of copies of the angle-angular momentum pair. We will return to this topic in more detail in a more technical work \cite{BBK}.

Finally, we feel obliged to point out that in the later publications by Aharonov {\em et al.} \cite{AR2005,Tollaksen2010}, their heuristic uncertainty relation\cite{APP} is instead presented as an actual inequality of the form \[\sdev{Q_T,\psi}\;\sdev{\MOD{P},{\psi}}\geq 2\pi\] in units where $\hbar=1$. This is, of course, incorrect  as the claimed lower bound is easily violated. There is no indication that these authors are considering a non-standard definition of the standard deviation. On the contrary, the explicit definitions in Ref. \onlinecite{Tollaksen2010} indeed confirm that standard deviations are used. Furthermore, the operator used by Aharonov {\em et al.} to measure the fringe width appears to lack the shift performed in \eqref{Pmod}. It follows that the value assigned to the fringe width increases as the number of illuminated slits is increased, indicating that the relevant features of the fringe width are not captured. In contrast, our choice \eqref{Pmod} yields the expected asymptotic behavior. The example application in the following section demonstrates this.

\section{The case of uniformly illuminated apertures}\label{exps} For our detailed discussion we focus on the uncertainty relation \eqref{APP1}, which describes the tradeoff between the spatial localisation of a quantum state incident on a multislit aperture and its fringe width.

We consider states obtained as follows: A single illuminated slit is assumed to prepare a quantum state described by a rectangular function of slit width, while a general aperture yields a suitable superposition of those. As the eigenspaces of $Q_T$ are excluded -- analysis of single-slit states is beyond the scope of this approach -- we consider superposition states of $m$ coherently illuminated slits, where $m$ is an even positive integer. These quantum states are but a subset of the quantum states that a multislit aperture can prepare, but they describe the important cases of the double slit aperture -- illustrated in Fig.~\ref{F1} -- and the periodic aperture and have a structure that makes a discussion of the uncertainty relation \eqref{APP1} both simple and insightful. They are of the form 
\begin{equation}\label{spies}\begin{split}
	\psi_m(x)=\frac{1}{\sqrt{m}}\sum_{j=1}^{m/2}
	\Bigl[\mathrm{rec}&\left(x+(2j-1){T/2}\right)\\
	&+\mathrm{rec}\left(x-(2j-1){T/2}\right)\Bigr],
	\end{split}
\end{equation}
where the function $\mathrm{rec}(x)$ is of rectangular shape,
\begin{equation}\label{g}
	\mathrm{rec}(x)=\left\{\begin{matrix}
			1/\sqrt a &\text{for $x\in [-a/2,a/2]$} \\
			0 &\text{for $x\notin[-a/2,a/2]$}
		\end{matrix}\right..
\end{equation}
The rectangular function is a popular choice for describing the profile across a single slit, although one may argue that it might not be the most physical choice. For the uncertainty product \eqref{APP1} this choice is actually entirely irrelevant as will become evident shortly. 

The momentum space wavefunction $\ft{\psi}_m(k)$ (the Fourier transform of $\psi_m$) is given by
\begin{equation}\label{sf}
	\ft{\psi}_m(k)=\sqrt{\frac{2a}{m\pi}}\;\sinc{\frac{a}{2}k}\sum_{j=1}^{m/2}{\Cos{\bracks{2j-1}\frac{T}{2}k}}.
\end{equation}

\begin{figure}
	\includegraphics[width=0.49\textwidth]{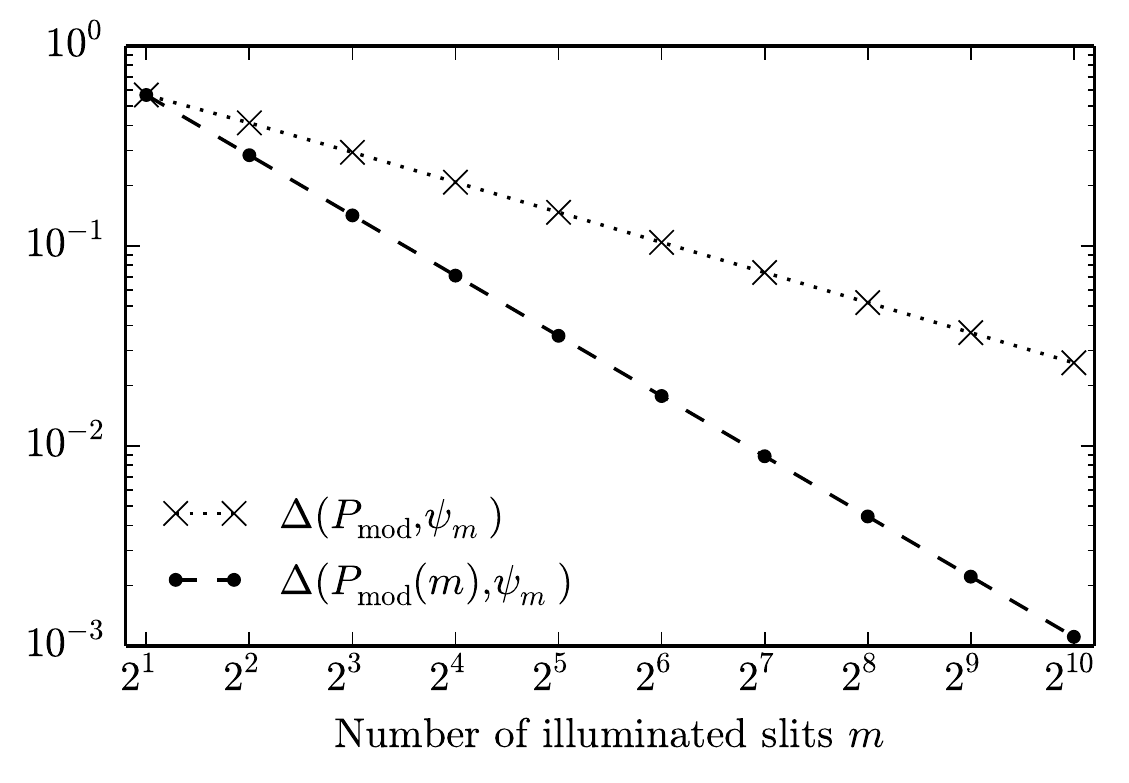}
	\caption{The slit number $m$ versus the standard deviation of $\MOD{P}$ in state $\psi_m$ is depicted (crosses on dotted line) for $T=5$, and also $\MOD{P}(m)$ (dots on dashed line), which is discussed in Sec.~\ref{Part2}.}
	\label{F5}
\end{figure}

For future reference, we define
\begin{equation}\label{fm}
	f_{m}(\kappa)=\sum_{j=1}^{m/2}{\Cos{\bracks{2j-1}\kappa}},
\end{equation}
where we have introduced the more natural variable $\kappa={Tk}/{2}$, proving particularly useful for integrations.

The standard deviation $\sdev{Q_T,\psi_m}$ is easy to compute analytically; one obtains\cite{Gneiting2011}
\begin{equation}\label{pos}
	\sdev{Q_T,\psi_m}=\frac{T}{2}\sqrt{\frac{m^2-1}{3}}.
\end{equation} For uniformly illuminated apertures the standard deviation of $Q_T$ increases linearly with the number of illuminated slits $m$. However, $\sdev{\MOD{P},{\psi_m}}$ requires some technical effort. Numerically computed values of $\sdev{\MOD{P},{\psi}_m}$ for $T=5$ are depicted in Fig.~\ref{F5} (crosses on dotted line), we deduce that this is approximately described by
\begin{equation}\label{naiveDeltaP}
	\sdev{\MOD{P},{\psi}_m}\approx\frac{0.32}{\sqrt{m}}.
\end{equation}
This result regarding the asymptotic behaviour of $\sdev{\MOD{P}}$ is confirmed analytically by a calculation in Appendix B, using a simplification that is discussed below and explicitly shown to hold in Appendix A.
We conclude immediately that the asymptotic behavior of the uncertainty product is divergent:
\begin{equation}\label{limit1}
	\lim_{m\rightarrow\infty}\sdev{Q_T,\psi_m}\;\sdev{\MOD{P},{\psi}_m}\propto\lim_{m\rightarrow\infty}\sqrt{m}=\infty\end{equation}
Note that for slit number $m$ (even) the value of $\sdev{\MOD{P},{\psi}_m}$ is independent of $a$. While this result might be expected by way of physical considerations, we show in Appendix B that it indeed follows when computing $\sdev{\MOD{P},{\psi}_m}$. This greatly simplifies any such computations. The calculation provided in Appendix B is very detailed; suffice it to say here that the result follows because $f_m(k)^2=f_m(k+jK)^2$ (with integer $j$) and because of a result on infinite sums in Ref. \onlinecite{BK48}.

\begin{align}
	\sdev{\MOD{P},{\psi}_m}&=\bracks{\Int{-\infty}{\infty}{{\MOD{P}(k)^2\,\psi_m(k)^2}}{k}}^{1/2}\nonumber
	\\&=\bracks{\frac{2^4}{T^2m\pi}{\Int{-\frac{\pi}{2}}{\frac{\pi}{2}}{\kappa^2\,f_{m}(\kappa)^2}{\kappa}}}^{1/2}\label{appA}
\end{align}
We proceed to calculate explicitly
\begin{align}\label{mom}
	\sdev{\MOD{P},{\psi}_2}&=\bracks{\frac{2^3}{T^2\pi}\Int{-\frac{\pi}{2}}{\frac{\pi}{2}}{\kappa^2\Cos{\kappa}^2}{\kappa}}^{\frac{1}{2}}\nonumber
	\\&=\frac{1}{T}\sqrt{\frac{\pi^2-6}{3}}
\end{align}
The uncertainty product for $\psi_2$ can be computed immediately from \eqref{pos} and \eqref{mom}:
\begin{align}
	\sdev{Q_T,\psi_2}\;\sdev{\MOD{P},{\psi}_2}&=\frac{1}{2}\sqrt{\bracks{\pi^2-6}/3}\nonumber\\&\approx0.568
\end{align}
This is surprisingly close to the lower bound already, and in fact equal to the value of the conventional uncertainty product \eqref{U} assigned to a particle confined to a box (in one dimension). This is not a coincidence; the central idea of the formulation of uncertainty discussed here is that a direct sum of such `boxes' is considered. More explicitly, for the particle in a box the spatial wavefunction of the lowest energy eigenstate is described by a half cosine pulse, whereas a single fringe of the double-slit state is described by a half cosine pulse. In these two considerations position space and momentum space are reversed: a spatial wavefunction is (typically) considered for the particle in the box, whereas a wavefunction in momentum space is considered here.

Any attempt at quantifying uncertainty using the uncertainty relation claimed in the work of Aharonov and Rohrlich \cite{AR2005} or Aharonov {\em et al.} \cite{Tollaksen2010} quickly leads to contradictions. The double-slit state $\psi_2$ (among others) violates the lower bound claimed there. Furthermore, the definition of $\MOD{P}$ implicit in Aharonov {\em et al.} \cite{APP,AR2005,Tollaksen2010} features an unsuitable choice of origin, leading to $\sdev{\MOD{P},\psi_m}$ increasing with increasing $m$; this is addressed here by means of the shift in \eqref{Pmod}.

In the following two sections material is developed that allows a more apt application of the uncertainty relation \eqref{APP1} through better adaptation to the specific structure of the quantum states $\psi_m$.

\section{Fine structure and refined $\MOD{P}$}

\subsection{The product form of $\ft{\psi}_{2^d}(k)$}\label{Interlude1} As an alternative to \eqref{sf}, for slit numbers that are powers of $2$, the wavefunctions $\ft{\psi}_{2^d}(k)$ can identically be described by 
\begin{equation}\label{pf}
	\ft{\psi}_{2^d}(k)=\sqrt{2^{d-1}\frac{a}{\pi}}\,\sinc{\frac{a}{2}\,k}\prod_{j=0}^{d-1}{\Cos{2^j\frac{T}{2}\,k}}
\end{equation}
The equivalence with \eqref{sf} is shown in Appendix C. How the product form arises conceptually is addressed in Appendix D. Observe that the product involves $d$ factors only and that for $d=0$ the correct expression for the single-slit wavefunction is obtained (whereas \eqref{sf} does not simplify to the single-slit state and consists of $2^{d-1}$ terms).
We proceed to discuss the structure highlighted by expressing wavefunctions in the product form of Eq.~\eqref{pf}, and then adapt our measure of the fringe width accordingly.
\subsection{Spread and fine structure by means of the product form}\label{Interlude2} 
\begin{figure*}\centering
\includegraphics[width=1\textwidth]{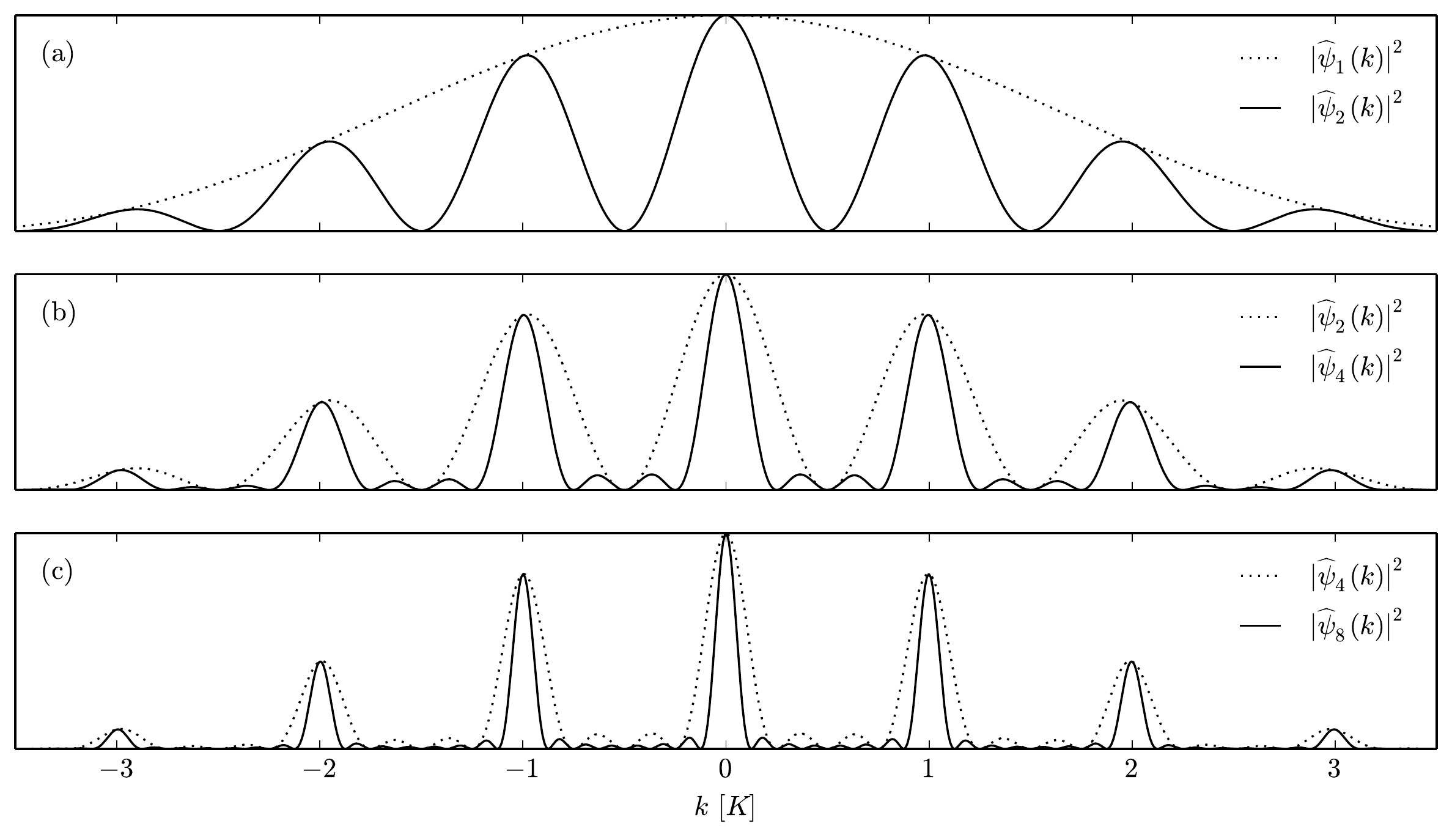}
\caption{The interference patterns associated with $\psi_2$, $\psi_4$ and $\psi_8$ are depicted; their recursive relationship is illustrated. Panel (a) shows $\modsquare{\ft{\psi}_2(k)}$, the first interference state and the {\em sinc} envelope (dotted line). Panel (b) shows $\modsquare{\ft{\psi}_4(k)}$ (solid line); $\modsquare{\ft{\psi}_2(k)}$  (dotted) serves as the envelope to $\modsquare{\ft{\psi}_4(k)}$. Panel (c) shows $\modsquare{\ft{\psi}_8(k)}$ (solid line); $\modsquare{\ft{\psi}_4(k)}$  (dotted) serves as the envelope to $\modsquare{\ft{\psi}_8(k)}$. Note that the depicted states are not normalized relative to each other so as to better demonstrate the shape and recursive relationship of the depicted states.}
\label{F3}
\end{figure*}

For the double-slit interference state the structure of the wavefunction in momentum space, given in \eqref{ds} and illustrated in Fig.~\ref{F1} (b), is easily read. There is a {\em sinc} envelope and {\em cosine} fringes. For this state the sum and the product form of the momentum wavefunction coincide. For larger $d$, however, the sum of {\em cosines} in \eqref{sf} obscures the structure of the respective state, making it impossible to tell the function determining the envelope from the one determining the fine structure. The product form, on the other hand, makes it very easy to distinguish the envelope from the fine structure. Note the following recursive relationship among the wavefunctions,
\begin{equation}\label{rec}
	\ft{\psi}_{2^{d+1}}(k)\propto\ft{\psi}_{2^d}(k)\Cos{2^d\frac{T}{2}\,k},
\end{equation}
implicit in \eqref{pf}.
This relationship is illustrated in Fig.~\ref{F3} for $\psi_{1}$, $\psi_{2}$, $\psi_{4}$ and $\psi_{8}$, i.e. for $d=0,1,2,3$. We can immediately draw conclusions about the spread of the interference pattern and the fringe width using the recursive relationship of \eqref{rec}.

Regarding the spread of the interference pattern, notice the following: The eight-slit wavefunction $\ft{\psi}_{8}(k)$ depicted in Fig.~\ref{F3} (c) (solid line) is contained in an envelope $\ft{\psi}_{4}(k)$ (dotted). Equally, the four-slit wavefunction $\ft{\psi}_{4}(k)$ depicted in Fig.~\ref{F3} (b) (solid line) is contained in an envelope $\ft{\psi}_{2}(k)$ (dotted). This argument applies recursively: the spread of $\ft{\psi}_{2^d}(k)$ is determined by $\ft{\psi}_{2^{d-1}}(k)$, and in turn the spread of $\ft{\psi}_{2^{d-1}}(k)$ is determined by $\ft{\psi}_{2^{d-2}}(k)$, all the way up to the single slit state $\ft{\psi}_{2^0}(k)$. This is an excellent illustration that the spread of the interference pattern is independent of the number of illuminated slits; it instead depends on the slit shape. The uncertainty product should not depend on the spread of the interference pattern nor, consequently, on the slit shape or width.

Considering the fine structure we note that while the {\em sinc} is common to the single-slit wavefunction $\ft{\psi}_1$ and the double-slit wavefunction $\ft{\psi}_{2}(k)$,  the latter also possesses fine structure described by the {\em cosine}. Doubling the number of illuminated slits (from $1$ to $2$) results in a {\em cosine} of frequency $T/2$. Doubling the number of illuminated slits once more (from $2$ to $4$) results in a {\em cosine} of double the frequency, i.e. $2(T/2)$, and hence a fringe width that is reduced by a factor of $2$. We observe that our initial fringe measure depends on the square root of the number of illuminated slits; see Eq.~\eqref{naiveDeltaP}. Yet we saw in this section that doubling the number of illuminated slits leads to a fine structure with doubled frequency; whatever measure might be used, this should be reflected. In the next section we discuss a modification that yields precisely the right asymptotic behavior.

\subsection{The refined modular momentum $\MOD{P}$}\label{Part2} We present here an application of \eqref{APP1} to multislit interferometry that takes into account the insights into the structure of interference wavefunctions as presented in the previous subsection. This entails an adaptation of the operator $\MOD{P}$ to the experimental setup, i.e. the number of illuminated slits $m$, which corresponds to an adaptation to the minimal period of the nodes occurring in the respective interference pattern. While the definition of $K$ in \eqref{K} is sufficient for the double-slit state $\psi_2$, we define now for general (even) $m$
\begin{equation}\label{K'}
	K'=2\pi/(nT)=4\pi/(mT)\,,
\end{equation}
as this reflects the behavior we observed in the previous section. We proceed to define a new operator adapted to the minimal period of the nodes
\begin{equation}
	P_{\mathrm{mod}}(m)=\bracks{P+K'/2\mod{K'}}-K'/2\,.\label{Pmod'}
\end{equation}
It is important to note that the middle expression in \eqref{K'} ensures that $\MOD{Q}$ commutes with the $\MOD{P}(m)$, so that we again obtain the uncertainty relation. The derivation of Sec.~\ref{APPsec} is extended to this more general case at the discretion of the reader.

The definition of $\MOD{P}(m)$ leads to a calculation very similar to the one performed in \eqref{appA}, but yields a result that depends inversely on $m$:
\begin{align}\label{mom'}
	\sdev{\MOD{P}{(m)},{\psi}_m}&=\bracks{\frac{2^2m}{T^2\pi}\Int{-\frac{\pi}{m}}{\frac{\pi}{m}}{\kappa^2\Cos{\frac{m}{2}\kappa}^2}{\kappa}}^{\frac{1}{2}}\nonumber\\
&=\bracks{\frac{2^5}{T^2\pi m^2}\Int{-\frac{\pi}{2}}{\frac{\pi}{2}}{y^2\Cos{y}^2}{y}}^{\frac{1}{2}}\nonumber
	\\&=\frac{2}{m T}\sqrt{\frac{\pi^2-6}{3}}\,.
\end{align}
\begin{figure}\centering
	\includegraphics[width=0.5\textwidth]{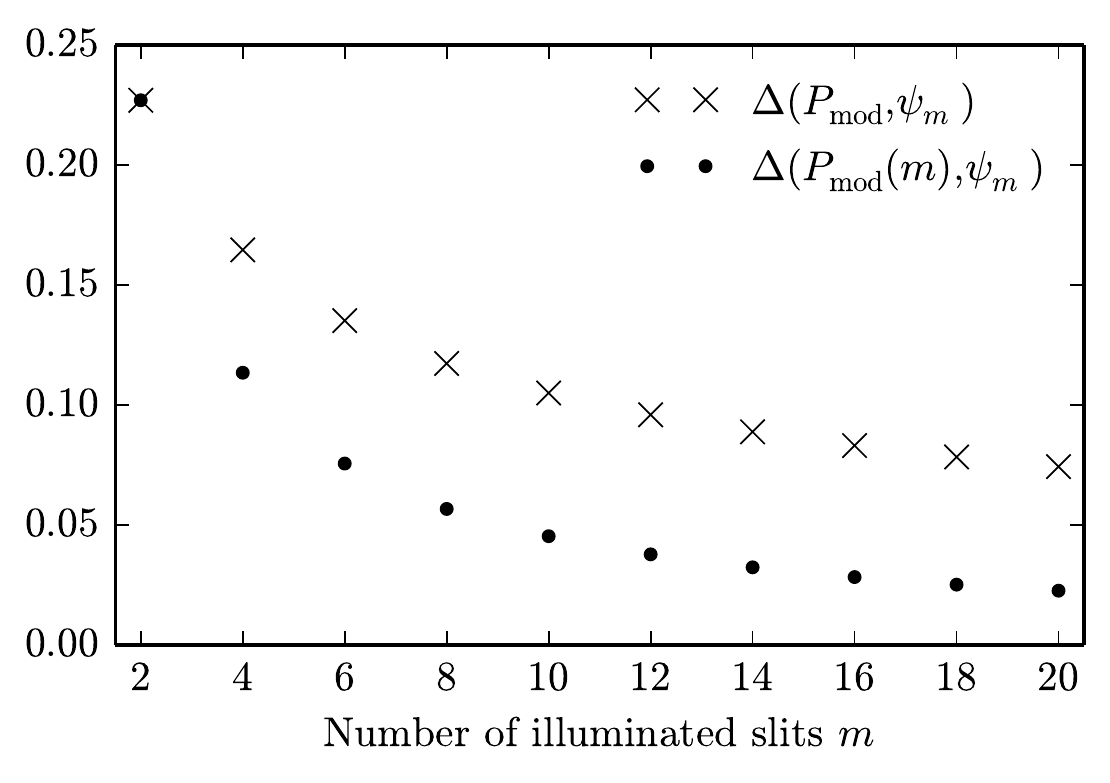}
	\caption{Values of fringe width associated with uniformly illuminated apertures versus the number of illuminated slits $m$, for $T=5$. The values assigned by $\sdev{\MOD{P},{\psi}_m}$ are represented using crosses, whereas using dots we indicate the values of $\sdev{\MOD{P}{(m)},{\psi}_m}$. Compare Fig.~\ref{F5}.}
	\label{F4}
\end{figure}
Physical considerations lead us to conclude the first line, mathematically this is shown in Appendix E.
The second line is obtained from the first line using the substitution $y=m\kappa/2$; note in particular how it compares to Eq.~\eqref{appA} where we found dependence on $m^{-1/2}$. The behavior of $\MOD{P}(m)$ is depicted in Fig.~\ref{F5} (dots on dashed line) and in Fig.~\ref{F4} (dots). 

\begin{figure}\centering
	\includegraphics[width=0.5\textwidth]{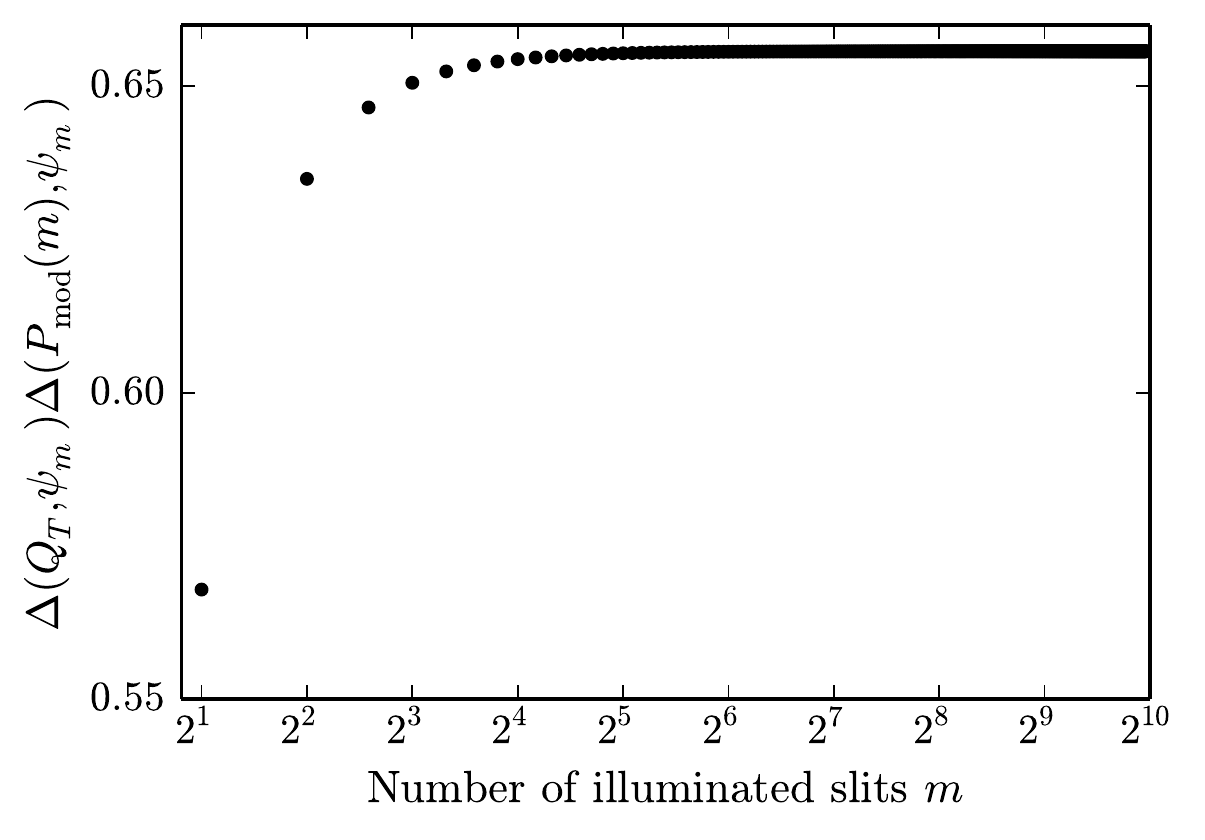}
	\caption{Convergence of the uncertainty product \eqref{limit2} for uniformly illuminated apertures of even slit number.}
	\label{d6}
\end{figure}

We can now take the limit to large $m$ for $\sdev{\MOD{P}{(m)},{\psi}_m}$, i.e. increasing the aperture while also adapting the operators $\MOD{P}(m)$. The resulting asymptotic behavior of the uncertainty product \eqref{APP1} is obtained immediately from \eqref{pos} and \eqref{mom'}
\begin{align}\label{limit2}
	 \lim_{m\rightarrow\infty}\sdev{Q_T,\psi_m}\sdev{\MOD{P}{(m)},{\psi}_m}&=\frac{1}{3}\sqrt{\pi^2-6}\nonumber\\&\approx0.656\,.
\end{align}
The convergent behavior is illustrated in Fig.~\ref{d6}.
We find that a finite value is obtained, while previously the uncertainty product \eqref{limit1} would diverge. This is the result of adapting the operator that measures the fringes to the interference setup considered by using the known relationship between the number of illuminated slits and the periodicity of the nodes discussed in the previous section. While the discussion in the previous section was restricted to slit numbers corresponding to powers of $2$ only, the expression for $\sdev{\MOD{P}{(m)},{\psi}_m}$ holds for all even $m$ as is illustrated in Fig.~\ref{F4}.

\section{Conclusion} A successful adaptation of the uncertainty relation \eqref{U} to multislit interferometry was presented. Based on an idea of Aharonov, Pendleton, Peterson \cite{APP}, we showed that a pair of uncertainty relations more suitable to multislit interferometry may be obtained by means of a suitable decomposition of the position and momentum observables. These uncertainty relations employ standard deviations, yet express the complementarity of spatial localisation and fringe width by virtue of the observables involved. We showed how these relations can be obtained with particular focus on the relevant commutation relations. Special care was taken to point out issues arising from domain questions and necessary boundary conditions.

We discussed in detail a certain subset of superposition states that can be expressed in a product form. The structure of interference wavefunctions in product form is easily read, which allows identifying the functions determining the spread of the interference pattern and the functions determining its fringe width. In particular, we discussed a recursive relationship between the interference wavefunctions that lead to a refined decomposition of the operator used to describe the fringe width, and to a finite uncertainty product for any (even) number of uniformly illuminated slits.

We argued that the formulation of uncertainty presented here is superior to the so-called Heisenberg uncertainty relation in the interferometric context. In particular, the measure of fringe width employed indeed gives reasonable (finite) values and shows the expected asymptotic behavior. Furthermore, the analysis does not depend on the slit width at all: the spatial localisation as well as the fringe width are independent of the value of the slit width. This is as it should be, as was pointed out. The new operators, however, do not address all problems. We pointed out the most important deficiencies.

Finally, we discussed a simple method for calculating the standard deviation of the operator used to express the fringe width. This simplification arises from the fact that the envelope has no effect on the calculation, and hence a single fringe becomes sufficient for a quantification of the fine structure.

We emphasize once more that the uncertainty relations discussed here are limited to quantum states that vanish periodically and are not suitable for a description of single-slit quantum states.

\section*{Acknowledgements} We used Enthought Canopy 1.3, Enthought, Inc. (\text{http://www.enthought.com}) provided through an academic license to perform the numerical calculations referenced and to create the associated figures.

\section*{Appendix A: Computing $\sdev{\MOD{P},{\psi}_m}$} 

Here we show that for slit number $m$ (even) the value of $\sdev{\MOD{P},{\psi}_m}$ is independent of $a$:
\begin{align*}
	&\sdev{\MOD{P},{\psi}_m}^2={\Int{-\infty}{\infty}{{\MOD{P}(k)^2\,\psi_m(k)^2}}{k}}
	\\
	&\qquad\qquad=\frac{2^4a}{T^3m\pi}\Int{-\infty}{\infty}{\MOD{P}(\kappa)^2\,\sinc{\frac{a}{T}\kappa}^2\,f_{m}(\kappa)^2}{\kappa},
\end{align*}
where we are using the shorthand $f_m(\kappa)$, which was introduced in \eqref{fm}. This integral may be decomposed into an infinite sum of integrals over the finite interval $K$,
\begin{align*}
=\frac{2^4a}{T^3m\pi}\sum_{j=-\infty}^{\infty}{\Int{\bracks{j-\tfrac{1}{2}}\pi}{\bracks{j+\tfrac{1}{2}}\pi}{(\kappa-j\pi)^2\,\sinc{\frac{a}{T}\kappa}^2\,f_{m}(\kappa)^2}{\kappa}}.
\end{align*}

We now substitute $u=\kappa-j\pi$ and immediately exploit $f_{m}(\kappa+j\pi)^2=f_{m}(\kappa)^2$,
\begin{align*}
	=\frac{2^4a}{T^3m\pi}{\Int{-\frac{\pi}{2}}{\frac{\pi}{2}}{u^2\,f_{m}(u)^2\sum_{j=-\infty}^{\infty}\sinc{\frac{a}{T}(u+j\pi)}^2}{u}}.
\end{align*}
The value of the series is known to be $T/a$ (see Eq.~(11) of Ref. \onlinecite{BK48}  and the derivation provided in Ref. \onlinecite{BP73}; note, however, that there is a factor of $1/\alpha$ missing in both the integral term and the series in Eq.~(1) of Ref. \onlinecite{BP73}); hence 
\begin{align*}
	\sdev{\MOD{P},{\psi}_m}^2&=\frac{2^4a}{T^3m\pi}\frac{T}{a}{\Int{-\frac{\pi}{2}}{\frac{\pi}{2}}{u^2\,f_{m}(u)^2}{u}}
	\\&=\frac{2^4}{T^2m\pi}{\Int{-\frac{K}{2}}{\frac{K}{2}}{u^2\,f_{m}(u)^2}{u}}.
\end{align*}The final expression is indeed equal to \eqref{appA}. 

\section*{Appendix B: $\sdev{\MOD{P},{\psi}_m}\sim 1/\sqrt{m}$} We show here that $\sdev{\MOD{P},{\psi}_m}$ in \eqref{naiveDeltaP} indeed asymptotically goes as $1/\sqrt{m}$. We use the result of Appendix A in order to simplify the necessary integration, which is the same as in \eqref{mom}, but for general $m$:
\begin{align}
	\sdev{\MOD{P},{\psi}_m}^2&=\frac{4}{mK}\bracks{\frac{2}{T}}^3{\Int{-\frac{\pi}{2}}{\frac{\pi}{2}}{\kappa^2\,f_{m}(\kappa)^2}{\kappa}}\label{resB}
	\\&=\frac{16}{m\pi T^{2}}\Int{-\frac{\pi}{2}}{\frac{\pi}{2}}{\kappa^2\bracks{\sum_{j=1}^{m/2}{\cos{(2j-1)\kappa}}}^2}{\kappa}.\nonumber
\end{align}
Note that
\begin{align}
	\bracks{\sum_{j=1}^{m/2}{\cos{(2j-1)\kappa}}}^2
	&=\frac{1}{4}\bracks{\frac{\sin{m\kappa}}{\sin{\kappa}}}^2\\
	&=\frac{1}{4}\frac{1-\cos{2m\kappa}}{1-\cos{2\kappa}}\,.\label{eq58}
\end{align}The standard deviation of the first term can be computed analytically
\begin{equation}
	\Int{-\pi/2}{\pi/2}{\frac{\kappa^2}{1-\cos{2\kappa}}}{\kappa}=\pi \ln{2}\,,
\end{equation}
whereas the second term yields
\begin{equation}
	\lim_{m\rightarrow\infty}\Int{-\frac{\pi}{2}}{\frac{\pi}{2}}{\frac{\kappa^2}{1-\cos{2\kappa}}\cos{2m\kappa}}{\kappa}=0\,,
\end{equation}
by the Riemann-Lebesgue lemma.
 Hence, we obtain
\begin{equation}\label{asymP}
	\sdev{\MOD{P},{\psi}_m}\approx\frac{2\sqrt{\ln{2}}}{T}\frac{1}{\sqrt{m}}\text{ for large }m,
\end{equation}
analytically confirming the asymptotic behaviour that was suggested by the numerical investigation, which is depicted in figure \ref{F5}, that led to \eqref{naiveDeltaP}. 

\section*{Appendix C: The Induction}
The equivalence of the summed interference wavefunction \eqref{sf} and the product form \eqref{pf} can be shown using mathematical induction. Starting the induction at $d=2$ (it holds trivially for $d=1$ and $d=0$),
\begin{align}
	\prod_{j=0}^{2-1}{\Cos{2^j\kappa}}&=\Cos{2^0\kappa}\Cos{2^1\kappa}\label{step1}
	\\&=2\sbracks{\Cos{(2-1){\kappa}}+\Cos{(2+1)\kappa}}\label{step2}
	\\&=2\sbracks{\Cos{{\kappa}}+\Cos{3\kappa}}
	\\&=2\sum_{j=1}^{2^{2-1}}{\Cos{(2j-1)\kappa}}
\end{align}
Hence for $d=2$ the product form equals the summation form. The known trigonometric identity \begin{equation}\label{TrigReduce}
	2\Cos{A}\Cos{B}=\Cos{A-B}+\Cos{A+B}
\end{equation}
was used going from \eqref{step1} to \eqref{step2}.
Now, assuming that the equivalence holds for the case $d$, i.e.
\begin{equation}\label{Assumption}
	2^{d-1}\prod_{j=0}^{d-1}{\Cos{2^j\kappa}}=\sum_{j=1}^{2^{d-1}}{\Cos{(2j-1)\kappa}}
\end{equation}
it is shown that the equivalence holds for the case $d+1$.
\begin{widetext}
\begin{align}
	2^{d}\prod_{j=0}^{d}{\Cos{2^j\kappa}}&=2\cdot2^{d-1}\prod_{j=0}^{d-1}{\Cos{2^j\kappa}}\cdot\Cos{2^d\kappa}
	\\&=2\cdot\sbracks{\sum_{j=1}^{2^{d-1}}{\Cos{(2j-1)\kappa}}}\cdot\Cos{2^d\kappa}
	\\&=\Cos{\bracks{2^d-1}\kappa}+\Cos{\bracks{2^d+1}\kappa}+\cdots+\Cos{\kappa}+\Cos{\bracks{2^{d+1}-1}\kappa}
	\\&=\Cos{\kappa}+\cdots+\Cos{\bracks{2^d-1}\kappa}+\Cos{\bracks{2^d+1}\kappa}+\cdots+\Cos{\bracks{2^{d+1}-1}\kappa}
	\\&=\sum_{j=1}^{2^{d}}{\Cos{(2j-1)\kappa}}
\end{align}
\end{widetext}
Hence it follows that
\begin{equation}\label{Res}
	2^{d-1}\prod_{i=1}^{d}\Cos{2^{i-1}}=\sum_{j=1}^{2^{d-1}}{\Cos{(2j-1)\kappa}}
\end{equation} which concludes the proof.

Note that the RHS of \eqref{Res} can be viewed as a Fourier series of the periodic function on the LHS. This Fourier series has the special property that its coefficients are either $1$ or $0$.

\section*{Appendix D: Deducing the product form} The alternative product form can be obtained easily by performing the Fourier transform differently. It is also possible to illustrate this difference in a diagram, as is done below. The different mathematical expressions leading to the sum or the product form are easily identified. Using an example of a uniformly illuminated aperture with eight slits, the Fourier transform is performed by considering each slit location as a delta function $\delta$, and the aperture as a sum of such. The following expression is found to express the  structure of the aperture
\begin{align*}
	f_{8}(x)=&\sbracks{\delta\bracks{x+1\,\frac{T}{2}}+\delta\bracks{x-1\,\frac{T}{2}}}
	\\&+\sbracks{\delta\bracks{x+3\,\frac{T}{2}}+\delta\bracks{x-3\,\frac{T}{2}}}
	\\&+\sbracks{\delta\bracks{x+5\,\frac{T}{2}}+\delta\bracks{x-5\,\frac{T}{2}}}
	\\&+\sbracks{\delta\bracks{x+7\,\frac{T}{2}}+\delta\bracks{x-7\,\frac{T}{2}}}
\end{align*}
See Eq.~\eqref{spies}, where this is the underlying structure. The Fourier transform is easily computed, the Fourier transform of a sum is the sum of Fourier transforms. The result is
\begin{align}
	\ft{f}_8(k)=&\Cos{\frac{T}{2}\,k}+\Cos{3\,\frac{T}{2}\,k}+\Cos{5\,\frac{T}{2}\,k}\nonumber
	\\&+\Cos{7\,\frac{T}{2}\,k}
\end{align}
Alternatively, the aperture may be described by the following convolutions
\begin{align*}
	g_8(x)=&\sbracks{\delta\bracks{x+1\,\frac{T}{2}}+\delta\bracks{x-1\,\frac{T}{2}}}
	\\&\ast\sbracks{\delta\bracks{x+2\,\frac{T}{2}}+\delta\bracks{x-2\,\frac{T}{2}}}
	\\&\ast\sbracks{\delta\bracks{x+4\,\frac{T}{2}}+\delta\bracks{x-4\,\frac{T}{2}}}
\end{align*}
The Fourier transform is also computed easily, noting that the Fourier transform of a convolution is a product of Fourier transforms; it leads immediately to the promised product form:
\begin{equation}
	\ft{g}_8(k)=\Cos{1\,\frac{T}{2}\,k}\cdot\Cos{2\,\frac{T}{2}\,k}\cdot\Cos{4\,\frac{T}{2}\,k}
\end{equation}
\begin{figure}
\includegraphics[width=0.49\textwidth]{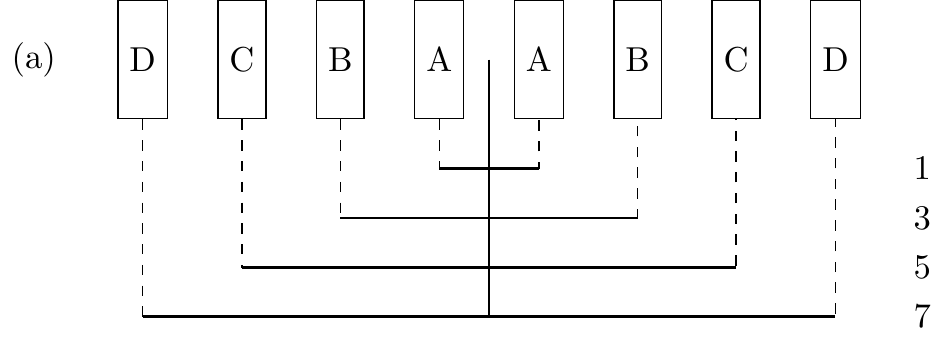}
\includegraphics[width=0.49\textwidth]{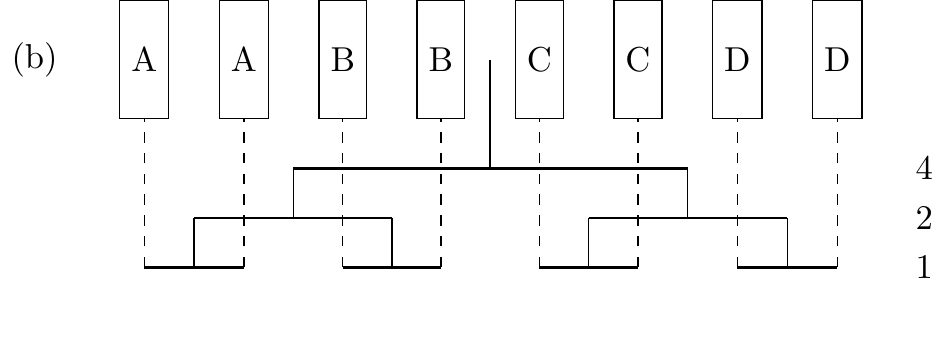}
	\caption{The two different ways of arranging the slits of an aperture in pairs are depicted for an aperture of eight slits. In (a), the usual way of pairing up slits across the origin is shown; this results in a sum. The numbers indicate the distance between the two members of the pair, they are in units of `slit separations'. In (b) an alternative way of pairing the slits is shown, which underlies the product form. The numbers indicate the distance of the successive pairings in units of `slit separations', down to the unit of a single pair of slits.}
	\label{Fig1}
\end{figure}

The observing reader may have noticed already that when computing the Fourier transform of a multislit aperture, there are two distinct ways of pairing up slits. The common way of pairing up two slits is easily identified: two slits with equal distance to the origin are considered a pair and their complex phases combined into a real cosine. This way of pairing up slits is illustrated in Fig.~\ref{Fig1} (a), the four pairs in the depicted 8-slit setup are denoted A, B, C and D. Compare \eqref{spies}, which explicitly highlights this structure.

The procedure underlying the product form is different, it is illustrated in Fig.~\ref{Fig1} (b). It entails successively dividing the aperture into halves, and the halves into quarters and so on, until pairs are left. In the illustrated example, dividing the initial aperture into halves results in a cosine factor scaled with the centre-to-centre distance of $4$ (in units of `half slit separations') between the halves. Dividing each of the two halves results in quarters gives a cosine factor scaled by $2$, which is the centre-to-centre distance between the quarters. Finally, the pairs of slits yield a cosine factor scaled with unity. Observe that the centre-to-centre distances indicated on the right-hand side times the number of occurrences is constant, e.g. on the lowest level $4$ pairs with a distance of unity are obtained whereas on the highest level there is one division with a centre-to-centre distance of $4$. This generalizes trivially to larger interference setups.

\section*{Appendix E: Computing $\sdev{\MOD{P}(m),\psi_{m}}$} The calculation here demonstrates that the calculation of $\sdev{\MOD{P}(m),\psi_{m}}$ can be simplified further than the result of Appendix A. This is done easily for the special case $m=2^d$, because this particular choice allows us to exploit the introduced product form. Below, a sketch of the general proof is included; the calculation turns out very similar but is substantially more tedious \cite{B14}. We start with an expression similar to \eqref{resB} but using $\MOD{P}(m)$ instead of $\MOD{P}$.
\begin{align*}
	\sdev{\MOD{P},{\psi}_m}^2=\frac{4}{mK}\bracks{\frac{2}{T}}^3{\Int{-\frac{\pi}{2}}{\frac{\pi}{2}}{\MOD{P}(m,\kappa)^2f_{m}(\kappa)^2}{\kappa}}
\end{align*}
We restrict ourselves to $m=2^d$ and substitute a product expansion
\begin{align*}
	f_{2^d}(\kappa)=2^{d-1}\prod_{j=0}^{d-1}\Cos{2^{j}\kappa}
\end{align*}
in place of the sum, giving
\begin{align*}
	=\frac{2^5}{T^3mK}\bracks{\frac{m}{2}}^2{\Int{-\frac{\pi}{2}}{\frac{\pi}{2}}{\MOD{P}(m,\kappa)^2\prod_{j=0}^{d-1}\Cos{2^{j}\kappa}^2}{\kappa}}
\end{align*}
Using the identity $\Cos{x}^2=(1+\cos{2x})/2$, it follows that each of the cosines from $j=0$ to $j=d-2$ contributes a factor of $1/2$, because the integrations are computed over multiples of the periods of the respective cosines. We proceed explicitly with the case $j=0$
\begin{align*}
	&=\frac{2^{3}m}{T^3K}\frac{1}{2}{\Int{-\frac{\pi}{2}}{\frac{\pi}{2}}{\MOD{P}(m,\kappa)^2\bracks{1+\cos{2\kappa}}\prod_{j=1}^{d-1}{\Cos{2^j\kappa}^2}}{\kappa}}
	\\&=\frac{2^{3}m}{T^3K}\frac{1}{2}{\Int{-\frac{\pi}{2}}{\frac{\pi}{2}}{\MOD{P}(m,\kappa)^2\prod_{j=1}^{d-1}{\Cos{2^j\kappa}^2}}{\kappa}}
\end{align*}
Repeating this another $d-2$ times contributes $(1/2)^{d-2}$
\begin{align*}
	=\frac{2^{3}m}{T^3K}{\frac{2}{m}}{\Int{-\frac{\pi}{2}}{\frac{\pi}{2}}{\MOD{P}(m,\kappa)^2\;{\Cos{2^{d-1}\kappa}^2}}{\kappa}}
\end{align*}
We immediately exploit the periodicity of the resulting function
\begin{align*}
	&=\frac{2^{3}m}{T^3K}\frac{2}{m}\frac{m}{2}{\Int{-\frac{\pi}{m}}{\frac{\pi}{m}}{\MOD{P}(m,\kappa)^2\;{\Cos{\frac{m}{2}\kappa}^2}}{\kappa}}
	\\&=\frac{2^{2}m}{T^2\pi}{\Int{-\frac{\pi}{m}}{\frac{\pi}{m}}{\kappa^2\;{\Cos{\frac{m}{2}\kappa}^2}}{\kappa}}
\end{align*}
The final expression is indeed equal to \eqref{mom'}, although here we only proved the special case $m=2^d$. In order to show this for all even $m$, two cases are required to be treated separately
\begin{equation*}
m/2=\left\{ \begin{matrix}
w &\text{if $m/2$ is even} \\
v &\text{if $m/2$ is odd}
\end{matrix}\right..
\end{equation*}
Using the following expressions
\begin{align*}
	&f_{2w}(\kappa)=2\Cos{\frac{m}{2}\kappa}\sum_{j=1}^{w/2}\Cos{(2j-1)\kappa}
	\\&f_{2v}(k)=\Cos{\frac{m}{2}\kappa}\sbracks{1+2\sum_{j=1}^{(v-1)/2}\Cos{2j\kappa}}
\end{align*}
and noting that the cross terms resulting from squaring $f_m(k)$ integrate to zero, these two cases can be calculated using the same trigonometric identity that was used above to reduce the power of a squared cosine.

\end{document}